\documentclass[
reprint,
superscriptaddress,
amsmath,
amssymb,
aps,
pre,
floatfix]{revtex4-2}
\usepackage{graphicx}
\usepackage{dcolumn}
\usepackage{bm}
\usepackage{physics,mathrsfs}
\usepackage{appendix}
\usepackage{comment}
\usepackage{placeins}
\usepackage{subfigure}
\usepackage{svg}
\usepackage{todonotes}
\usepackage{CJKutf8}
\usepackage{csquotes}
\usepackage{bbm}
\usepackage{dsfont}
\usepackage{hyperref}
\usepackage[capitalise]{cleveref}
\DeclareUnicodeCharacter{2009}{\,}

\usepackage{xcolor}
\begin{document}

\preprint{APS/123-QED}

\title{Thermodynamically Consistent Lindbladians for Quantum Stochastic Thermodynamics}

\author{Jin-Fu Chen (\begin{CJK*}{UTF8}{gbsn}陈劲夫\end{CJK*})}
\email{jinfuchen@lorentz.leidenuniv.nl}
\affiliation{Instituut-Lorentz, Universiteit Leiden, P.O. Box 9506, 2300 RA Leiden, The Netherlands}
\affiliation{$\langle aQa^L\rangle$ Applied Quantum Algorithms, Universiteit Leiden, The Netherlands}

\date{\today}
\begin{abstract}
We develop a thermodynamically consistent Lindblad approach for quantum stochastic thermodynamics, where each bath dissipator satisfies the local quantum detailed balance condition. We formulate the thermodynamic geometry to determine the optimal protocol in the slow-driving regime. Thermodynamic consistency leads to the joint fluctuation theorem for quantum work and heat, with the corresponding joint statistics obtained from a tilted master equation and generally described by a quasiprobability. We apply this framework to the quantum Brownian motion master equation, refining it to satisfy detailed balance, deriving optimal protocols at different temperatures and friction coefficients, and studying the work statistics. Our approach thus enables the study of nonequilibrium thermodynamics in open quantum systems.
\end{abstract}

\maketitle

\section{Introduction}\label{sec:introduction}

Stochastic thermodynamics provides a simple yet effective framework for understanding nonequilibrium behaviors in microscopic systems. By modeling heat exchange with effective stochastic processes, it has produced insights for understanding the thermodynamic laws and reducing the irreversibility at the microscopic level, e.g., fluctuation theorems \cite{jarzynski_nonequilibrium_1997,crooks_entropy_1999,seifert_stochastic_2012} and optimal control \cite{crooks_measuring_2007,schmiedl_optimal_2007,blaber_optimal_2023}. The detailed balance condition ensures thermodynamic consistency \cite{soret_thermodynamic_2022}, so that work and heat in stochastic processes obey the thermodynamic laws. More recently, significant progress has been made in thermodynamic bounds \cite{ohga_thermodynamic_2023,kolchinsky_thermodynamic_2024,liang_thermodynamic_2024,owen_universal_2020,maes_frenetic_2017}, including thermodynamic uncertainty relations \cite{barato_thermodynamic_2015,horowitz_thermodynamic_2020,timpanaro_thermodynamic_2019,hasegawa_fluctuation_2019,van_vu_thermodynamic_2023}. These results deepen our understanding of nonequilibrium thermodynamics and provide guiding principles for controlling microscopic systems \cite{seifert_stochastic_2012}. For example, the power and efficiency of microscopic heat engines \cite{ma_universal_2018,liang_minimal_2025} and chemical motors \cite{zhai_power-efficiency_2025} are enhanced by reducing irreversibility and harnessing fluctuations.

Despite significant progress in classical systems, extending stochastic thermodynamics to quantum systems remains challenging. One major challenge lies in defining fluctuating quantum work and heat, which are typically formulated by the two-point measurement scheme \cite{talkner_fluctuation_2007,esposito_nonequilibrium_2009,campisi_colloquium_2011,strasberg_quantum_2021}. However, this scheme can alter the subsequent evolution of the system, as the initial measurement collapses the system into an energy eigenstate. A no-go theorem imposes constraints on defining work and heat in quantum systems \cite{perarnau-llobet_no-go_2017,hovhannisyan_energy_2024}, which motivates quasiprobability descriptions in quantum thermodynamics \cite{gherardini_quasiprobabilities_2024,francica_class_2022,solinas_full_2015,pei_exploring_2023,upadhyaya_non-abelian_2024,arvidsson-shukur_properties_2024}. 

Following the same spirit of stochastic thermodynamics, Lindblad master equations provide a natural description of quantum systems coupled to heat baths \cite{cavina_slow_2017,chetrite_quantum_2012,liu_characteristic_2016,de_chiara_quantum_2022,manzano_quantum_2018,ptaszynski_thermodynamics_2019,ostilli_thermalization_2017}. The quantum detailed balance condition \cite{alicki_detailed_1976,carlen_gradient_2017,fagnola_generators_2010,ramezani_quantum_2018,gilyen_quantum_2024,chen_efficient_2023,chen_boosting_2024} makes this description thermodynamically consistent. Reference~\cite{chen_boosting_2024} introduced a transformation that constructs finite-temperature Lindbladians satisfying this condition from their infinite-temperature counterparts. This method can be used to describe open quantum systems and was applied there to thermal-state preparation and spectral-gap optimization. Here, we use the same construction to study thermodynamic geometry and joint statistics of work and heat in driven nonequilibrium processes. These Lindbladians can include quantum jumps without a definite energy change, which have not been addressed in most previous studies of quantum stochastic thermodynamics \cite{liu_characteristic_2016,de_chiara_quantum_2022,manzano_quantum_2018,ptaszynski_thermodynamics_2019}.

In this work, we develop a framework for quantum stochastic thermodynamics to study nonequilibrium thermodynamics of open quantum systems in finite-time driving processes. We require that each dissipator in the Lindbladian satisfies the detailed balance condition with respect to the corresponding heat bath. For an initially thermal system, this thermodynamic consistency leads to the joint fluctuation theorem for quantum work and heat, while the first moments give the ensemble averages of work and heat.
For a suitable set of observables, the density-matrix dynamics reduces to equations for their expectation values. In the slow driving regime, these equations determine the thermodynamic metric and the optimal protocol of changing the control parameters.

Finally, we apply the framework to quantum Brownian motion \cite{caldeira_quantum_1983,breuer_theory_2002}. We refine the quantum Brownian motion master equation to satisfy the quantum detailed balance condition and remain completely positive at all temperatures. We obtain optimal protocols for varying the frequency of the harmonic potential. The high-temperature protocols recover known results from stochastic thermodynamics \cite{schmiedl_optimal_2007,dechant_underdamped_2017,chen_microscopic_2022}, while the low-temperature protocols are new. We also study the full counting statistics of work and verify the fluctuation theorem of work \cite{jarzynski_nonequilibrium_1997} and the quantum-classical correspondence principle \cite{jarzynski_quantum-classical_2015}.

The rest of the work is organized as follows. In Sec.~\ref{sec:detailed-balance}, we define the thermodynamically consistent dynamics. In Sec.~\ref{sec:effective}, we introduce the observable dynamics and the slow-driving expansion. In Sec.~\ref{sec:geometry}, we use this expansion to derive the thermodynamic geometry and the optimal protocols. In Sec.~\ref{sec:fcs}, we develop the tilted master equation for the joint statistics of work and heat. In Sec.~\ref{sec:qbm}, we study a quantum harmonic oscillator as an example.

\begin{figure}
    \centering
    \includegraphics[width=\linewidth]{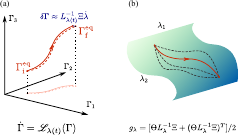}
    \caption{Diagram of our framework: (a) Evolution of observables $\Gamma$ in the quasistatic process (red dashed curve) and the finite-time driving process (red solid curve), with the lag in the finite-time driving process estimated by the first slow-driving correction in Eq.~\eqref{eq:slow_solution}. (b) The manifold of control parameter space under the thermodynamic length. The red solid curve shows the geodesic as the optimal protocol. }
    \label{fig:diagram}
\end{figure}

\section{Thermodynamically consistent Lindblad dynamics}\label{sec:detailed-balance}

We use Lindblad generators whose bath dissipators separately satisfy the quantum detailed balance condition \cite{fagnola_generators_2010,chen_efficient_2023,gilyen_quantum_2024,ramezani_quantum_2018}. This local detailed balance condition gives thermodynamically consistent driven dynamics when the system is coupled to multiple baths at different temperatures.

The evolution of an open quantum system is described by the Lindblad master equation $\dot{\boldsymbol{\rho}}=\mathcal{L}_{\lambda(t)}[\boldsymbol{\rho}]$ with the Lindbladian
\begin{align}
    \mathcal{L}_{\lambda(t)}[\bullet]=-i[\boldsymbol{h},\bullet]+\sum_k \mathcal{D}_k[\bullet],\label{eq:master_eq}
\end{align}
where the Hamiltonian $\boldsymbol{h}$ and the dissipators $\mathcal{D}_k$ are time-dependent due to the control parameters $\lambda(t)$. The heat baths are effectively described by the dissipators in the Lindblad form, with each $\mathcal{D}_k$ corresponding to the $k$th heat bath. 
For each bath, local detailed balance relates $\mathcal D_k$ to its adjoint through the system thermal state $\boldsymbol{\rho}_{\beta_k}=e^{-\beta_k \boldsymbol{h}}/Z_\lambda(\beta_k)$, where $\beta_k$ is the inverse temperature of bath $k$ and $Z_\lambda(\beta_k)=\mathrm{tr}[e^{-\beta_k\boldsymbol h(\lambda)}]$ \cite{breuer_theory_2002}. We suppress the control-parameter dependence of $\boldsymbol{\rho}_{\beta_k}$ and $\mathcal D_k$ below. The state $\boldsymbol{\rho}_{\beta_k}$ is a fixed point of that dissipator. We give the explicit construction below, following Ref.~\cite{chen_boosting_2024}.

\subsection{Construction of dissipators satisfying local detailed balance}

We first construct the dissipator at fixed control parameters and later consider its dependence on $\lambda(t)$ for driven dynamics. For bath $k$, define $\mathcal G_k[\bullet]=\boldsymbol{\rho}_{\beta_k}^{1/2}\bullet\boldsymbol{\rho}_{\beta_k}^{1/2}$. The Hilbert--Schmidt adjoint is defined by $\mathrm{tr}\{\boldsymbol X^\dagger\mathcal D_k[\boldsymbol Y]\}=\mathrm{tr}\{(\mathcal D_k^\sharp[\boldsymbol X])^\dagger \boldsymbol Y\}$. Local detailed balance requires
\begin{align}
\mathcal D_k^\sharp[\bullet]=\mathcal G_k^{-1}\mathcal D_k\mathcal G_k[\bullet] .
\label{eq:local_qdb_map}
\end{align}
Equivalently, $\mathcal D_k^\sharp$ is self-adjoint under the Kubo--Martin--Schwinger (KMS) inner product $\langle \boldsymbol X,\boldsymbol Y\rangle_{\boldsymbol{\rho}_{\beta_k}}=\mathrm{tr}(\boldsymbol X^\dagger\boldsymbol{\rho}_{\beta_k}^{1/2}\boldsymbol Y\boldsymbol{\rho}_{\beta_k}^{1/2})$.

Following Ref.~\cite{chen_boosting_2024}, we start from an infinite-temperature dissipator. A tilde denotes an operator or superoperator at infinite temperature:
\begin{equation}
\tilde{\mathcal D}_k[\bullet]
=\sum_n\tilde\gamma_{k n}\left(
\tilde{\boldsymbol A}_{k n}\bullet\tilde{\boldsymbol A}_{k n}^\dagger
-\frac{1}{2}\{\tilde{\boldsymbol A}_{k n}^\dagger\tilde{\boldsymbol A}_{k n},\bullet\}\right).
\end{equation}
Here $n$ labels the jump channels and $\tilde\gamma_{kn}\geq0$ is the corresponding rate.
Each jump $\tilde{\boldsymbol A}_{k n}$ is paired with its adjoint at the same rate. The generator is unital and Hilbert--Schmidt self-adjoint, as required by Eq.~\eqref{eq:local_qdb_map} at $\beta_k=0$.

At finite temperature, the Lindblad operators are constructed as
\begin{align}
\boldsymbol A_{k n}=\boldsymbol{\rho}_{\beta_k}^{1/4}\tilde{\boldsymbol A}_{k n}\boldsymbol{\rho}_{\beta_k}^{-1/4}
=e^{-\beta_k \boldsymbol h/4}\tilde{\boldsymbol A}_{k n}e^{\beta_k \boldsymbol h/4}.
\label{eq:finite_temperature_jump}
\end{align}
Equation~\eqref{eq:finite_temperature_jump} introduces the thermal bias between forward and reverse transitions. In the energy basis, a matrix element connecting $|a\rangle$ to $|b\rangle$ is multiplied by $e^{-\beta_k(E_b-E_a)/4}$. Thus, when $E_b>E_a$, the transition rate from $|a\rangle$ to $|b\rangle$ is smaller than the reverse rate by the Boltzmann factor $e^{-\beta_k(E_b-E_a)}$. Moreover, if $\tilde{\boldsymbol A}_{k n'}=\tilde{\boldsymbol A}_{k n}^\dagger$, the transformed pair obeys $\boldsymbol{\rho}_{\beta_k}^{-1/2}\boldsymbol A_{k n}\boldsymbol{\rho}_{\beta_k}^{1/2}=\boldsymbol A_{k n'}^\dagger$. This conjugation symmetry, together with the adjoint pairing, ensures that the jump term satisfies quantum detailed balance.

To ensure that the dissipator as a whole satisfies quantum detailed balance, we introduce a coherent part following Ref.~\cite{chen_efficient_2023}. Defining $\boldsymbol R_k=\sum_n\tilde\gamma_{k n}\boldsymbol A_{k n}^\dagger\boldsymbol A_{k n}$, the finite-temperature dissipator becomes
\begin{align}
\mathcal D_k[\bullet]=-i[\boldsymbol K_k,\bullet]
+\sum_n\tilde\gamma_{k n}\boldsymbol A_{k n}\bullet\boldsymbol A_{k n}^\dagger
-\frac{1}{2}\{\boldsymbol R_k,\bullet\}.
\label{eq:finite_temperature_dissipator}
\end{align}
Here $\boldsymbol K_k$ is a bath-induced coherent correction to ensure the quantum detailed balance condition for the dissipator. Substituting Eq.~\eqref{eq:finite_temperature_dissipator} into Eq.~\eqref{eq:local_qdb_map}, we obtain
\begin{align}
\{\boldsymbol{\rho}_{\beta_k}^{-1/2},\boldsymbol{\rho}_{\beta_k}^{1/4}\boldsymbol K_k\boldsymbol{\rho}_{\beta_k}^{1/4}\}
=\frac{i}{2}[\boldsymbol{\rho}_{\beta_k}^{-1/2},\boldsymbol{\rho}_{\beta_k}^{1/4}\boldsymbol R_k\boldsymbol{\rho}_{\beta_k}^{1/4}],
\label{eq:coherent_correction}
\end{align}
which can be solved in the energy basis. Let $\boldsymbol h|a\rangle=E_a|a\rangle$, with $\nu=E_b-E_a$ denoting an energy difference. Following Ref.~\cite{chen_efficient_2023}, the solution is
\begin{align}
\boldsymbol K_k&=\frac{i}{2}\sum_{\nu}\tanh\left(\frac{\beta_k\nu}{4}\right)\boldsymbol R_{k,\nu},\\
\boldsymbol R_{k,\nu}&:=\sum_{E_b-E_a=\nu}|b\rangle\langle b|\boldsymbol R_k|a\rangle\langle a|.
\label{eq:coherent_correction_energy_basis}
\end{align}
The components obey $\boldsymbol R_k=\sum_{\nu}\boldsymbol R_{k,\nu}$ and $[\boldsymbol h,\boldsymbol R_{k,\nu}]=\nu\boldsymbol R_{k,\nu}$. Thus, $\boldsymbol R_{k,\nu}$ collects the matrix elements associated with the energy difference $\nu$. Since $\boldsymbol R_{k,-\nu}=\boldsymbol R_{k,\nu}^\dagger$, $\boldsymbol K_k$ is Hermitian. For each component $\boldsymbol R_{k,\nu}$, the commutator generated by $\boldsymbol K_k$ cancels the deviation from detailed balance in the anticommutator term. Thus, the complete dissipator satisfies Eq.~\eqref{eq:local_qdb_map}. If $[\boldsymbol R_k,\boldsymbol h]=0$, only the $\nu=0$ component remains and $\boldsymbol K_k=0$. For quantum Brownian motion it is generally nonzero.

The dissipator in Eq.~\eqref{eq:finite_temperature_dissipator} generates completely positive, trace-preserving dynamics. Since the identity operator $\boldsymbol I$ satisfies $\mathcal D_k^\sharp[\boldsymbol I]=0$, Eq.~\eqref{eq:local_qdb_map} gives $\mathcal D_k[\boldsymbol{\rho}_{\beta_k}]=0$. Thus, the system thermal state associated with bath $k$ is a fixed point of the corresponding dissipator, and the master equation is thermodynamically consistent. When the system is coupled to several reservoirs at different temperatures, the fixed point of the full generator is generally a nonequilibrium steady state rather than a thermal state.

For a single bath, writing $\mathcal G=\mathcal G_1$, the complete generator obeys $\mathcal L^\sharp[\bullet]-\mathcal G^{-1}\mathcal L\mathcal G[\bullet]=2i[\boldsymbol h,\bullet]$. The fact that the thermal state is a fixed point under the same detailed balance condition is used in Appendix~\ref{sec:fcs-derivation} to verify the fluctuation relation. In Sec.~\ref{sec:qbm-generator}, Eqs.~\eqref{eq:finite_temperature_jump} and \eqref{eq:coherent_correction} are applied to $\tilde{\boldsymbol A}\propto x$, which gives Eqs.~\eqref{eq:A_quantum_Brownian} and \eqref{eq:K_quantum_Brownian}.

Our framework yields two key results in nonequilibrium thermodynamics of open quantum systems: the optimization of slow-driving control strategies through thermodynamic geometry and, subsequently, the full counting statistics of work and heat generated by the tilted master equation.

\section{Observable dynamics and the slow-driving expansion}\label{sec:effective}

Directly solving the evolution of the density matrix in nonequilibrium processes can be challenging. If a chosen operator subspace is invariant under the adjoint dynamics, the evolution can instead be written as a closed set of equations for expectation values. The dynamics after adding the counting fields can be treated in the same way when the corresponding tilted equations also close (see Fig.~\ref{fig:diagram}(a)). We use observables $\boldsymbol O_j$ to characterize the state through $\Gamma_j = \mathrm{tr}(\boldsymbol{\rho}\boldsymbol O_j)$ during the finite-time driving process. If the operator basis depends explicitly on time, an additional term $\mathrm{tr}(\boldsymbol\rho\,\partial_t\boldsymbol O_j)$ must be included. For the fixed basis used here, their evolution is
\begin{align}
\dot{\Gamma}=\mathscr{L}_{\lambda(t)}(\Gamma)=\mathscr{H}(\Gamma)+\sum_{k}\mathscr{D}_{k}(\Gamma),\label{eq:Gammadot}
\end{align}
where $\mathscr{H}(\Gamma)$ and $\mathscr{D}_k (\Gamma)$ represent the unitary evolution and the dissipation dynamics for the observables, respectively. 
Explicit differential equations can be determined for specific systems, such as a two-level system or a quantum harmonic oscillator. An exact dimensional reduction is obtained when the closed operator set is substantially smaller than the density-matrix representation. When exact closure is unavailable, a projected operator dynamics provides a systematic approximation. For the quantum Brownian model in a harmonic potential, the tilted equations close after the second moments are augmented by $\chi$, as shown in Sec.~\ref{sec:qbm-counting}.
Equation~\eqref{eq:Gammadot} closes if the adjoint Lindbladian maps the span of the chosen observables, including the identity, into itself. A sufficient linear condition is
\begin{align}
\mathcal{L}_{\lambda}^{\sharp}[\boldsymbol O_j]
=f_j(\lambda)\,\boldsymbol I+\sum_l L_{jl}(\lambda)\boldsymbol O_l.
\label{eq:observable_closure}
\end{align}
Here $j$ and $l$ label the chosen observables. Equation~\eqref{eq:observable_closure} gives $\dot{\Gamma}_j=f_j+\sum_lL_{jl}\Gamma_l$. When this condition is not satisfied by a small operator set, higher-order observables enter. Their dynamics can still be treated in a finite Krylov space, as discussed in Sec.~\ref{sec:qbm-extensions}. In the quadratic Brownian model, the Gaussian state preserves closure of the second moments.

In principle, the optimal control of $\lambda(t)$ can be formulated as a Pontryagin’s maximum principle problem \cite{pontryagin1962,boscain_introduction_2021,Cavina2018,chen_optimal_2024} under Eq.~\eqref{eq:Gammadot} with a cost function, such as the excess work, though it is challenging to solve. Fortunately, a geodesic approach optimizes the slow-driving protocol using the thermodynamic length \cite{sivak_thermodynamic_2012,scandi_thermodynamic_2019,chen_extrapolating_2021}. For slow driving, we expand the observables around the instantaneous steady state.

We restrict the following discussion to the linear form
\begin{equation}
\dot\Gamma=L_\lambda\Gamma+f_\lambda.
\label{eq:slow_rescaled}
\end{equation}
When Eq.~\eqref{eq:observable_closure} holds, Eq.~\eqref{eq:slow_rescaled} is the exact reduced dynamics, with $L_\lambda=(L_{jl})$ and $(f_\lambda)_j=f_j$. Nonlinear effective dynamics can arise when higher-order observables are approximated by functions of the selected observables. Such dynamics may be linearized about the instantaneous steady state in the slow-driving regime. Here we consider only the linear dynamics in Eq.~\eqref{eq:slow_rescaled} for simplicity. Our approach can also be extended to more general nonlinear effective dynamics.

Consider a fixed path $\lambda(s)$, with $s=t/\tau$, where $\tau$ is the protocol duration. Here $\tau_{\mathrm{rel}}$ denotes a characteristic relaxation time. For Lindblad dynamics, it can be estimated from the spectral gap $\Delta_{\mathcal L}$, the smallest nonzero decay rate, as $\tau_{\mathrm{rel}}\sim\Delta_{\mathcal L}^{-1}$~\cite{chen_boosting_2024}. In the reduced dynamics of Eq.~\eqref{eq:slow_rescaled}, the same time scale can be obtained from the inverse magnitude of the real part of the nonzero eigenvalue of $L_\lambda$ closest to zero. If the decay rates vary along the path, $\tau_{\mathrm{rel}}$ is set by the slowest relaxation along that path. Defining the perturbation parameter $\epsilon=\tau_{\mathrm{rel}}/\tau$, we expand the solution as
\begin{equation}
\begin{aligned}
\Gamma(s;\epsilon)&=\sum_{n=0}^{\infty}\epsilon^n\Gamma^{(n)}(s),\\
\sum_{n=0}^{\infty}\epsilon^{n+1}\partial_s\Gamma^{(n)}
&=\tau_{\mathrm{rel}}L_\lambda
\sum_{n=0}^{\infty}\epsilon^n\Gamma^{(n)}
+\tau_{\mathrm{rel}}f_\lambda.
\end{aligned}
\label{eq:slow_series}
\end{equation}
We match equal powers of $\epsilon$ and obtain a hierarchy for the linear equation  
\begin{equation}
\begin{aligned}
0&=L_\lambda\Gamma^{(0)}+f_\lambda,\\
\partial_s\Gamma^{(n-1)}
&=\tau_{\mathrm{rel}}L_\lambda\Gamma^{(n)},\qquad n\geq1,\\
\Gamma^{(n)}&=\left(\tau_{\mathrm{rel}}L_\lambda\right)^{-1}
\partial_s\Gamma^{(n-1)}
=\left[\left(\tau_{\mathrm{rel}}L_\lambda\right)^{-1}
\partial_s\right]^n\Gamma^{(0)}.
\end{aligned}
\label{eq:slow_hierarchy}
\end{equation}
The factor $\epsilon$ multiplying the time derivative shifts the order by one. Thus, the derivative of $\Gamma^{(n-1)}$ on the left is matched with $\Gamma^{(n)}$ on the right. The inverse is taken on the decaying subspace. Apart from an initial relaxation period, the recursion gives
\begin{equation}
\begin{aligned}
\Gamma
&=\Gamma^{(0)}+\sum_{n=1}^{\infty}\epsilon^n
\left[\left(\tau_{\mathrm{rel}}L_\lambda\right)^{-1}
\partial_s\right]^n\Gamma^{(0)}\\
&=\Gamma^{(0)}+L_\lambda^{-1}\dot\Gamma^{(0)}
+L_\lambda^{-1}\frac{d}{dt}
\left(L_\lambda^{-1}\dot\Gamma^{(0)}\right)+\cdots.
\end{aligned}
\label{eq:slow_solution}
\end{equation}
Here $\partial_s$ acts on everything to its right, including $L_\lambda^{-1}$, and the second line is written in physical time $t$. When Eq.~\eqref{eq:slow_rescaled} is exact, Eq.~\eqref{eq:slow_solution} gives the full slow-driving series. When Eq.~\eqref{eq:slow_rescaled} comes from linearizing a nonlinear dynamics, only its first correction is fixed at this level. In both cases, $\Gamma-\Gamma^{(0)}=L_\lambda^{-1}\dot\Gamma^{(0)}+O(\epsilon^2)$, which is the only term needed for the thermodynamic metric below.
Below, we focus on optimizing the control of open quantum systems coupled to a single heat bath.

\section{Thermodynamic geometry and optimal control}\label{sec:geometry}

We consider a system coupled to a single heat bath. At fixed control parameters, the Lindbladian drives the system to the thermal state $\boldsymbol{\rho}_\beta$, and the quasistatic observables take their equilibrium values, $\Gamma^{(0)}=\Gamma^\mathrm{eq}$. For a finite-time driving process, irreversibility is quantified by the excess work $W^\mathrm{ex}=W-W^\mathrm{eq}$, with equilibrium work given by the free energy change, $W^\mathrm{eq}=\int\mathrm{tr}(\boldsymbol{\rho}_\beta d\boldsymbol h)=\Delta F$.

We minimize the excess work over protocols $\lambda(t)$ with fixed endpoints $\lambda(0)=\lambda_{\mathrm{i}}$ and $\lambda(\tau)=\lambda_{\mathrm{f}}$, and with a fixed duration $\tau$:
\begin{align}
W^{\mathrm{ex}}[\lambda]
=\int_0^\tau dt\,\dot{\lambda}^{\top}\Theta
\bigl(\Gamma-\Gamma^{\mathrm{eq}}\bigr).
\label{eq:control_problem}
\end{align}
The geodesic solves this problem to leading order in slow driving.

We now connect the first slow-driving correction in Eq.~\eqref{eq:slow_solution} to the excess work. The Hamiltonian changes as $\partial_t\boldsymbol h=\sum_{ij}\dot\lambda_i\Theta_{ij}\boldsymbol O_j$, where $\Theta$ is the coupling matrix between the control parameters and observables and $\lambda=(\lambda_1,\lambda_2,\ldots,\lambda_n)^{\top}$. The excess-work rate is
\begin{equation}
\dot W^{\mathrm{ex}}
=\dot\lambda^{\top}\Theta
(\Gamma-\Gamma^{\mathrm{eq}}).
\end{equation}
For a single bath, $\Gamma^{(0)}=\Gamma^{\mathrm{eq}}$. Its change along the protocol is $\dot\Gamma^{\mathrm{eq}}=\Xi\dot\lambda$, with $\Xi_{ji}=\partial_{\lambda_i}\Gamma_j^{\mathrm{eq}}$. The first correction in Eq.~\eqref{eq:slow_solution} then gives $\Gamma-\Gamma^{\mathrm{eq}}=L_\lambda^{-1}\Xi\dot\lambda$. Substitution into the excess-work rate yields
\begin{equation}
\begin{aligned}
\dot W^{\mathrm{ex}}
&\approx\dot\lambda^{\top}
\Theta L_\lambda^{-1}\Xi\dot\lambda
=\dot\lambda^{\top}g_\lambda\dot\lambda,\\
g_\lambda
&=\frac{1}{2}\left[
\Theta L_\lambda^{-1}\Xi
+(\Theta L_\lambda^{-1}\Xi)^{\top}
\right].
\end{aligned}
\label{eq:excess_work_dot}
\end{equation}
Only the symmetric part contributes to the quadratic form, which gives the second equality. We use the term thermodynamic metric on the control subspace where $g_\lambda$ is positive semidefinite. It then defines the thermodynamic length along a path \cite{sivak_thermodynamic_2012,scandi_thermodynamic_2019,chen_extrapolating_2021},
\begin{equation}
\mathfrak{L}
=\int_0^\tau dt\,
\sqrt{\dot\lambda^{\top}g_\lambda\dot\lambda}
=\int_\mathcal P\sqrt{d\lambda^{\top}g_\lambda d\lambda}.
\end{equation}
Here $\mathcal P$ is the path traced by $\lambda(t)$ in the control-parameter space.
For fixed duration, the Cauchy--Schwarz inequality gives
\begin{equation}
\begin{aligned}
\mathfrak{L}^2
&\leq \tau\int_0^\tau dt\,
\dot\lambda^{\top}g_\lambda\dot\lambda
=\tau W^{\mathrm{ex}},\\
W^{\mathrm{ex}}&\geq\frac{\mathfrak{L}^2}{\tau}.
\end{aligned}
\label{eq:length_bound}
\end{equation}
Equality holds at this order for constant thermodynamic speed. Minimizing the path length between the fixed endpoints gives the geodesic and hence the leading slow-driving optimal protocol shown in Fig.~\ref{fig:diagram}(b).
The thermodynamic metric describes the leading excess work and determines the slow-driving optimal protocols. Work and heat fluctuations require introducing counting fields in the master equation, which we introduce next.

\section{Full counting statistics from the tilted master equation}\label{sec:fcs}

Full counting statistics describe the fluctuations of work and heat in a nonequilibrium process. In classical stochastic thermodynamics, both quantities are random variables \cite{seifert_stochastic_2012}. In quantum systems, however, work and heat are not observables \cite{talkner_fluctuation_2007}. Coherence and noncommuting operators then make their joint distribution a quasiprobability in general \cite{gherardini_quasiprobabilities_2024,solinas_full_2015,francica_class_2022,pei_exploring_2023,upadhyaya_non-abelian_2024,arvidsson-shukur_properties_2024}.

The joint statistics of work and heat are encoded in the generating function $\chi(u,\{v_k\})$. Unless the counted contributions admit a compatible measurement interpretation, $\chi(u,\{v_k\})$ need not be the Laplace transform of a positive measure. The integral fluctuation relation constrains this generating function, while its first derivatives give the ensemble averages of work and heat. In commuting or classical limits, the quasiprobability recovers an ordinary probability distribution.

In the Lindblad approach, the heat exchange is effectively described by the dissipator $\mathcal{D}_k$. This allows introducing counting fields in the master equation to evaluate the full counting statistics of quantum work and heat (for closed quantum systems, see Ref.~\cite{fei_quantum_2018}). We evaluate the generating function with a quantum Feynman--Kac-type tilted equation. Related tilted evolutions for classical master equations and diffusion processes are given in Refs.~\cite{jarzynski_master_1997,hummer_free_2001}. Here the construction includes several reservoirs and counts work and heat simultaneously. The quantum work is counted by adding the counting field $u$ to the generator,
\begin{align}
\mathcal{W}_{u}[\bullet]=\frac{\partial e^{\frac{u\boldsymbol h}{2}}}{\partial t}e^{-\frac{u\boldsymbol h}{2}}\bullet+\bullet e^{-\frac{u\boldsymbol h}{2}}\frac{\partial e^{\frac{u\boldsymbol h}{2}}}{\partial t}.\label{eq:Wurho}
\end{align}
We incorporate the counting field for heat exchange with the $k$th bath in the dissipator as 
\begin{align}
\mathcal{D}_{k,v_{k}}[\bullet]=e^{\frac{v_{k}\boldsymbol h}{2}}\mathcal{D}_{k}[e^{-\frac{v_{k}\boldsymbol h}{2}}\bullet e^{-\frac{v_{k}\boldsymbol h}{2}}]e^{\frac{v_{k}\boldsymbol h}{2}}.\label{eq:Dvrho}
\end{align}
Let $\boldsymbol\eta(u,\{v_k\};t)$ denote the tilted density operator, with the initial condition $\boldsymbol\eta(u,\{v_k\};0)=\boldsymbol\rho(0)$. Its Feynman--Kac equation is
\begin{align}
\dot{\boldsymbol{\eta}}=-i[\boldsymbol h,\boldsymbol{\eta}]+\mathcal{W}_u[\boldsymbol{\eta}]+\sum_k\mathcal{D}_{k,v_k}[\boldsymbol{\eta}],
\label{eq:eta_dot}
\end{align}
which gives a tilted generator with the counting fields $u$ and $v_k$.
We define the time-dependent generating function by $\chi(u,\{v_k\};t)=\mathrm{tr}[\boldsymbol\eta(u,\{v_k\};t)]$, and use $\chi(u,\{v_k\})\equiv\chi(u,\{v_k\};\tau)$ for the complete protocol. Formally, $\chi(u,\{v_k\})=\langle e^{u w+\sum_kv_kq_k}\rangle$, where $w$ is work done on the system and $q_k$ is heat transferred from bath $k$ to the system. The brackets denote an average over the joint quasiprobability; they do not assume a positive distribution. The first moments are $W=\langle w\rangle$ and $Q_k=\langle q_k\rangle$.

As a consequence of local detailed balance, the joint fluctuation theorem for quantum work and heat holds for the initial equilibrium condition specified below. We take $\boldsymbol\rho(0)=e^{-\beta_S\boldsymbol h(0)}/Z_\mathrm{i}(\beta_S)$, where $\beta_S$ is the initial inverse temperature of the system and $Z_\mathrm{i}(\beta_S)=\mathrm{tr}[e^{-\beta_S\boldsymbol h(0)}]$. We also define $Z_\mathrm{f}(\beta_S)=\mathrm{tr}[e^{-\beta_S\boldsymbol h(\tau)}]$ and $\Delta F_S=-\beta_S^{-1}\ln[Z_\mathrm{f}(\beta_S)/Z_\mathrm{i}(\beta_S)]$. We couple the system to multiple heat baths during a finite-time driving process. Setting $u=-\beta_S$ and $v_k=-(\beta_S-\beta_k )$, we find that $\boldsymbol{\eta}(t)=e^{-\beta_S \boldsymbol h(t)}/Z_\mathrm{i}(\beta_S)$ solves the tilted master equation and gives
\begin{align}
    \left\langle e^{-\beta_S w+\sum_k (\beta_k-\beta_S) q_k}\right\rangle=e^{-\beta_S \Delta F_S}.\label{eq:FT_work_heat}
\end{align}
The derivation is given in Appendix~\ref{sec:fcs-derivation}. For a single heat bath ($k=1$) with an initial inverse temperature $\beta_S = \beta_1 = \beta$, the work fluctuation theorem $\left\langle e^{-\beta w}\right\rangle = e^{-\beta \Delta F}$ \cite{jarzynski_nonequilibrium_1997} holds without the information of heat exchange. For relaxation processes without external driving, the fluctuation theorem for heat exchange, $\left\langle e^{\sum_{k}(\beta_{k}-\beta_{S})q_{k}}\right\rangle =1$, is recovered \cite{jarzynski_classical_2004}.

The moments follow from derivatives of $\chi(u,\{v_k\})$ at zero counting fields. In particular, $W=\left.\partial_u\chi\right|_{u=0,\,v_k=0}$ and $Q_k=\left.\partial_{v_k}\chi\right|_{u=0,\,v_k=0}$. Expanding $\boldsymbol{\eta}$ to first order shows that these expressions agree with $W=\int_0^\tau dt\,\mathrm{tr}(\boldsymbol{\rho}\,\partial_t\boldsymbol h)$ and $Q_k=\int_0^\tau dt\,\mathrm{tr}(\boldsymbol h\mathcal D_k[\boldsymbol{\rho}])$ \cite{alicki_quantum_1979,quan_quantum_2007}. Higher derivatives give higher moments.

\section{Quantum Brownian motion}\label{sec:qbm}

Brownian motion serves as a fundamental example in stochastic thermodynamics. Yet, the nonequilibrium thermodynamics of its quantum counterpart, the quantum Brownian motion, remains less explored. To the best of our knowledge, previous studies have focused on work statistics in dragged oscillators \cite{funo_path_2018} and heat statistics in relaxation dynamics \cite{chen_quantumclassical_2021}. These results alone do not provide the driven work and heat statistics needed to analyze closed cycles of quantum heat engines.

\subsection{Thermodynamically consistent generator}\label{sec:qbm-generator}
To investigate the nonequilibrium thermodynamics of quantum Brownian motion under time-dependent frequency, we apply our framework and derive the refined quantum Brownian Lindbladian
$    \mathcal{L}_\omega[\bullet]=-i[\boldsymbol h,\bullet]+\mathcal{D}[\bullet]$. The Hamiltonian of a quantum harmonic oscillator is $\boldsymbol h=p^2/(2m)+m\omega^2 x^2 /2$. Here $x$ and $p$ are the canonical position and momentum operators, $[x,p]=i$, $m>0$ is the mass, and $\omega>0$ is the oscillator frequency. We always use $\hbar=k_{\mathrm B}=1$ for simplicity. We write the dissipator in the same form as Eq.~\eqref{eq:finite_temperature_dissipator}, $\mathcal{D}[\bullet]=-i[\boldsymbol K,\bullet]+(\kappa/2)\boldsymbol A\bullet\boldsymbol A^\dagger-\{\boldsymbol R,\bullet\}/2$, with $\boldsymbol R=(\kappa/2)\boldsymbol A^\dagger\boldsymbol A$. Thus, both $\boldsymbol R$ and $\boldsymbol K$ include the friction coefficient $\kappa>0$.
The quantum detailed balance condition holds by choosing the coherent correction and Lindblad operator as  
\begin{align}
\boldsymbol K&=\frac{\kappa}{4\cosh(\beta\omega/2)}(xp+px),\label{eq:K_quantum_Brownian}\\\boldsymbol A&=\sqrt{m\omega\coth(\frac{\beta\omega}{4})}x+i\sqrt{\frac{\tanh(\beta\omega/4)}{m\omega}}p.\label{eq:A_quantum_Brownian}
\end{align}
These expressions follow directly from the general construction above. The resulting Lindbladian is completely positive and trace preserving, satisfies detailed balance, and has the thermal state $\boldsymbol{\rho}_\beta$ as its steady state.

For the oscillator, $a$ is the annihilation operator, $x=(a+a^\dagger)/\sqrt{2m\omega}$, and $[\boldsymbol h,a]=-\omega a$. 
Before applying the finite-temperature transformation, we may choose a seed operator $\tilde{\boldsymbol A}\propto x$ and the rate $\tilde\gamma=\kappa/2$, as in the Brownian master equation \cite{caldeira_quantum_1983,breuer_theory_2002}. The corresponding infinite-temperature dissipator is Hilbert--Schmidt self-adjoint, while the overall normalization of the seed operator is fixed by the classical limit. 
Its moment equations reduce to the classical damping term $-2\kappa\Gamma_{p^2}$ and the noise term $2\kappa m/\beta$ in the limit $\beta\omega\to0$, as shown in Sec.~\ref{sec:qbm-gaussian}.

\subsection{Closed Gaussian dynamics and classical limit}\label{sec:qbm-gaussian}
We choose the effective parametrization $\Gamma=(\Gamma_{x^2},\Gamma_{xp},\Gamma_{p^2})^{\top}$ for the observables $x^2$, $(xp+px)/2$, and $p^2$. The evolution equation of observables is given by
\begin{align}
    \dot{\Gamma}=\mathscr{L}_{\omega}(\Gamma)&=L_{\omega}\Gamma+f,
    \label{eq:qbm-second-moments}
\end{align}
where the linear part and the noise term are 
\begin{align}
L_{\omega}&=\left(\begin{array}{@{}ccc@{}}
-\kappa[1-\operatorname{sech}(\frac{\beta\omega}{2})] & \frac{2}{m} & 0\\
-m\omega^{2} & -\kappa & \frac{1}{m}\\
0 & -2m\omega^{2} & -\kappa[1+\operatorname{sech}(\frac{\beta\omega}{2})]
\end{array}\right),\\
f&=\left(\begin{array}{c}
\dfrac{\kappa}{2m\omega}\tanh(\dfrac{\beta\omega}{4})\\[3pt]
0\\[3pt]
\dfrac{\kappa m\omega}{2}\coth(\dfrac{\beta\omega}{4})
\end{array}\right).
\end{align}
This evolution equation allows evaluating the excess work $W^\mathrm{ex}$ in finite-time driving processes with time-dependent frequency $\omega(t)$. 
The steady-state values of the observables are solved from $\mathscr{L}_\omega (\Gamma)=0$, yielding $\Gamma^\mathrm{eq}_{x^2}=\coth(\beta\omega/2)/(2m\omega)$, $\Gamma^\mathrm{eq}_{xp}=0$, and $\Gamma_{p^2}^{\mathrm{eq}}=m\omega\coth(\beta\omega/2)/2$, corresponding to the thermal state of the harmonic oscillator.
The derivation of Eq.~\eqref{eq:qbm-second-moments} from the Brownian master equation is given in Appendix~\ref{sec:qbm-second-moment-derivation}.

We check the validity of the above equation of motion by taking the classical limit $\beta \omega\rightarrow 0$, and we obtain
\begin{align}
    L_\omega&\longrightarrow\left(\begin{array}{ccc}
0 & \frac{2}{m} & 0\\
-m\omega^{2} & -\kappa & \frac{1}{m}\\
0 & -2m\omega^{2} & -2\kappa
\end{array}\right),\\
f&\longrightarrow\left(\begin{array}{c}
0\\
0\\
\frac{2\kappa m}{\beta}
\end{array}\right),
\end{align}
which recovers the result in Ref.~\cite{chen_optimal_2024}.

\subsection{Excess work, thermodynamic metric, and optimal protocols}\label{sec:qbm-optimal}

The average work of varying the frequency $\omega (t)$ is given by
\begin{align}
   W=\int_{0}^{\tau}dt\,\dot{\omega}\mathrm{tr}(\boldsymbol{\rho}\frac{\partial \boldsymbol h}{\partial\omega})=\int_{0}^{\tau}dt\,m\omega\dot{\omega}\Gamma_{x^2 },
\end{align}
and for a quasistatic process, the work is
\begin{align}
    W^{\mathrm{eq}}&=\int_{\omega_{0}}^{\omega_{\tau}}d\omega\, m\omega\Gamma_{x^{2}}^{\mathrm{eq}}
    =\frac{1}{\beta}\ln\left[\frac{\sinh(\frac{\beta\omega_{\tau}}{2})}{\sinh(\frac{\beta\omega_{0}}{2})}\right].
\end{align}

The excess-work rate is
\begin{align}
\dot W^{\mathrm{ex}}&=m\omega\dot{\omega}(\Gamma_{x^{2}}-\Gamma_{x^{2}}^{\mathrm{eq}}).
\end{align}
We obtain the coupling matrix $\Theta$ between the control parameters and observables and the equilibrium response $\Xi$ as
\begin{align}
   \Theta&=(\begin{array}{ccc}
m\omega & 0 & 0\end{array}),\\
\Xi&=\left(
\begin{array}{c}
 -\frac{(\beta  \omega +\sinh (\beta  \omega )) \operatorname{csch}^2\left(\frac{\beta  \omega }{2}\right)}{4 m \omega ^2} \\
 0 \\
 \frac{1}{4} m (\sinh (\beta  \omega )-\beta  \omega ) \operatorname{csch}^2\left(\frac{\beta  \omega }{2}\right) \\
\end{array}
\right).
\end{align}
\begin{widetext}
The rate of the excess work is
\begin{align}
\dot{W}^{\mathrm{ex}}\approx\frac{\operatorname{csch}^{2}(\frac{\beta\omega}{2})\left[\beta\omega\left(\kappa^{2}+4\omega^{2}\right)+\kappa^{2}\left(2\sinh(\frac{\beta\omega}{2})+\sinh(\beta\omega)+\beta\omega\operatorname{sech}(\frac{\beta\omega}{2})\right)\right]}{4\kappa\omega\left(\kappa^{2}+4\omega^{2}-\kappa^{2}\operatorname{sech}^{2}(\frac{\beta\omega}{2})\right)}\dot{\omega }^2.
\end{align}
In the slow driving regime, the optimal protocol has constant thermodynamic speed, $\dot W^{\mathrm{ex}}=C$, where $C>0$ is fixed by the endpoints and the duration $\tau$. Thus,
\begin{align}
\frac{\dot{\omega}}{\omega}=\sqrt{C\cdot\frac{2\kappa(\cosh(\beta\omega)-1)\left(\kappa^{2}+4\omega^{2}-\kappa^{2}\operatorname{sech}^{2}(\frac{\beta\omega}{2})\right)}{\omega\left[\beta\omega\left(\kappa^{2}+4\omega^{2}\right)+\kappa^{2}\left(2\sinh(\frac{\beta\omega}{2})+\sinh(\beta\omega)+\beta\omega\operatorname{sech}(\frac{\beta\omega}{2})\right)\right]}}.\label{eq:dotomegadividedomega}
\end{align}
Introducing the dimensionless frequency $z=\beta\omega$, the thermodynamic length is evaluated as
\begin{align}
\mathfrak{L}&=\int_{\beta\omega_{0}}^{\beta\omega_{\tau}}\sqrt{\frac{4z^{3}+\left[z+2\sinh(\frac{z}{2})+\sinh(z)+z\operatorname{sech}(\frac{z}{2})\right]\beta^{2}\kappa^{2}}{4 z\sinh^{2}(\frac{z}{2})\left[4z^{2}+\beta^{2}\kappa^{2}(1-\operatorname{sech}^{2}(\frac{z}{2}))\right]\beta\kappa}}dz.
\end{align}
\end{widetext}

The high-temperature limit $\beta\omega\rightarrow 0$ gives
\begin{align}
\dot{W}_{\beta\omega\rightarrow0}^{\mathrm{ex}}=\frac{\kappa^{2}+\omega^{2}}{\beta\kappa\omega^{4}}\dot{\omega}^{2}.
\end{align}
The optimal protocol becomes
\begin{align}
\frac{\dot{\omega}}{\omega}=\sqrt{C\cdot\frac{\beta\kappa\omega^{2}}{\kappa^{2}+\omega^{2}}}.\label{eq:dotomegadividedomegahighT}
\end{align}

For small $\kappa$, the excess-work rate reduces to
$\dot{W}_{\beta\omega\rightarrow0}^{\mathrm{ex}}=\dot{\omega}^{2}/(\beta\kappa\omega^{2})$. For large $\kappa$, it becomes
$\dot{W}_{\beta\omega\rightarrow0}^{\mathrm{ex}}=\kappa\dot{\omega}^{2}/(\beta\omega^{4})$. The corresponding protocols are listed in Table~\ref{tab:optimal_protocol}.

For large frequency or the low temperature limit $\beta\omega \rightarrow\infty$, the excess work is dominated by
\begin{align}
\dot{W}_{\beta\omega\rightarrow\infty}^{\mathrm{ex}}=\frac{\kappa}{2\omega\left(\kappa^{2}+4\omega^{2}\right)}\dot{\omega}^2.
\end{align}
The optimal protocol becomes
\begin{align}
\frac{\dot{\omega}}{\omega}=\sqrt{C\cdot\frac{2(\kappa^{2}+4\omega^{2})}{\kappa\omega}}.\label{eq:dotomegadividedomegalowT}
\end{align}

For small $\kappa$, the excess-work rate reduces to
$\dot{W}_{\beta\omega\rightarrow\infty}^{\mathrm{ex}}=\kappa\dot{\omega}^{2}/(8\omega^{3})$. For large $\kappa$, it becomes
$\dot{W}_{\beta\omega\rightarrow\infty}^{\mathrm{ex}}=\dot{\omega}^{2}/(2\kappa\omega)$. The corresponding protocols are again given in Table~\ref{tab:optimal_protocol}.

\begin{figure}
    \centering
    \includegraphics[width=\linewidth]{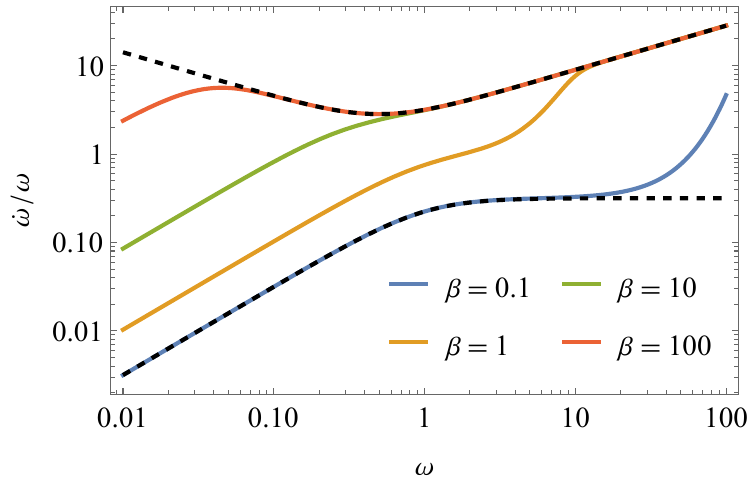}
    \caption{Optimal driving rate for varying $\omega$ from $0.01$ to $100$. We plot the velocity $\dot{\omega}/\omega$ from Eq.~\eqref{eq:dotomegadividedomega} as a function of $\omega$. We set $\kappa=1$ and $C=1$ and choose different inverse temperatures $\beta=0.1,1,10,100$. The lower black dashed curve shows the high-temperature result in Eq.~\eqref{eq:dotomegadividedomegahighT}, evaluated at $\beta=0.1$, while the upper black dashed curve shows the low-temperature result in Eq.~\eqref{eq:dotomegadividedomegalowT}.}
    \label{fig:optimal-protocol}
\end{figure}

We show the result in Fig.~\ref{fig:optimal-protocol}. We set $\kappa=1$ and $C=1$, and plot the optimal speed of varying the frequency $\dot{\omega}/\omega $ as a function of $\omega$ under different temperatures.

In the slow driving regime, we determine the thermodynamic length and the optimal protocol for varying the frequency $\omega$. Analytical results of the optimal protocols are obtained in different limits, the high (low) temperature limits $\beta\omega\rightarrow0$ ($\infty$), the underdamped (overdamped) limits $\kappa/\omega\rightarrow0$ ($\infty$), as summarized in Table~\ref{tab:optimal_protocol}, where $s=t/\tau$ denotes the rescaled time with $\tau$ the duration of the whole process. The high-temperature limit $\beta\omega\rightarrow0$ recovers the protocols obtained in classical Brownian motion \cite{schmiedl_optimal_2007,dechant_underdamped_2017,chen_microscopic_2022}. The low-temperature protocols describe the quantum regime $\beta\omega\rightarrow\infty$. 

\begin{table}
\caption{Optimal protocols for quantum Brownian motion in different limits.\label{tab:optimal_protocol}}
\begin{ruledtabular}
\begin{tabular}{lcc}
  & $\displaystyle\frac{\kappa}{\omega}\rightarrow0$ & $\displaystyle\frac{\kappa}{\omega}\rightarrow\infty$  \\
$\displaystyle\beta\omega\rightarrow 0$ & $\displaystyle\omega_{0}\left(\frac{\omega_{\tau}}{\omega_{0}}\right)^{s}$ & $\displaystyle\frac{\omega_{0}}{\left(\frac{\omega_{0}}{\omega_{\tau}}-1\right)s+1}$ \\
$\displaystyle\beta\omega\rightarrow \infty$ & $\displaystyle\frac{\omega_{0}}{\left[1+\left(\sqrt{\frac{\omega_{0}}{\omega_{\tau}}}-1\right)s\right]^{2}}$ & $\displaystyle\omega_{0}\left[1+\left(\sqrt{\frac{\omega_{\tau}}{\omega_{0}}}-1\right)s\right]^{2}$
\end{tabular}
\end{ruledtabular}
\end{table}

\subsection{Full counting statistics of work and numerical results}\label{sec:qbm-counting}

\begin{figure}
    \centering
    \includegraphics[width=\linewidth]{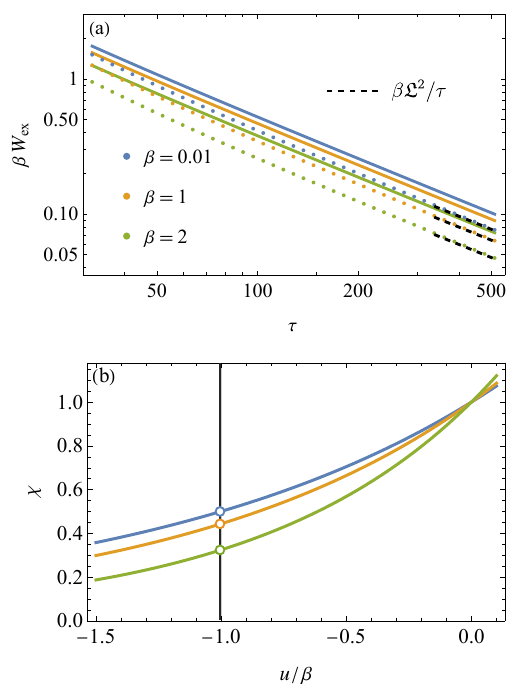}
    \caption{Results for nonequilibrium processes in the quantum Brownian motion model: (a) Scaled excess work $\beta W^\mathrm{ex}$ for compression processes ($\omega_0=0.2$ and $\omega_\tau =5$) at different inverse temperatures $\beta=0.01,1,2$. We compare the exponential protocol (curves) and the slow-driving optimal protocol (dots) for fixed initial and final frequencies. The black dashed lines show the leading prediction from thermodynamic length, $\beta\mathfrak{L}^2/\tau$.
    (b) The generating function of work for compression processes ($\omega_0=1$ and $\omega_\tau =2$) with the duration $\tau=50$. The solid curves are obtained by propagating the tilted dynamics. The open circles at $u/\beta=-1$ show the independently evaluated equilibrium values $e^{-\beta\Delta F}$. Their agreement with the curves verifies the work fluctuation theorem.}
    \label{fig:excesswork}
\end{figure}

For the Brownian model, the tilted evolution remains closed in the Gaussian sector. Starting from the thermal state, parity is preserved by the frequency driving and the tilted evolution, so the first moments remain zero. We use $\Gamma_u=(\chi,\Gamma_{x^2},\Gamma_{xp},\Gamma_{p^2})^{\top}$, with $\chi(u,t)=\mathrm{tr}[\boldsymbol\eta(u,t)]$. For $\boldsymbol O_j\in\{x^2,(xp+px)/2,p^2\}$, the normalized tilted moments are defined by $\Gamma_j(u,t)=\mathrm{tr}[\boldsymbol\eta(u,t)\boldsymbol O_j]/\chi(u,t)$. At $u=0$, $\chi=1$ and these quantities reduce to the physical moments in Sec.~\ref{sec:qbm-gaussian}. We extend $\mathscr L_\omega$ in Eq.~\eqref{eq:qbm-second-moments} to $\Gamma_u$ by assigning it a zero first component and letting it act on the last three components as before. Their evolution has the compact form
\begin{align}
\dot{\Gamma}_u=\mathscr{W}_{u}(\Gamma_u)+\mathscr{L}_\omega(\Gamma_u).
\label{eq:dotGamma}
\end{align}
The counting term $\mathscr{W}_u$ also determines $\dot\chi$. Its explicit form is given in Appendix~\ref{sec:qbm-counting-derivation}. At $u=0$, Eq.~\eqref{eq:dotGamma} reduces to the physical Brownian dynamics. At the end of the protocol, $u=-\beta$ gives $\chi(-\beta;\tau)=\sinh(\beta\omega_0/2)/\sinh(\beta\omega_\tau/2)=e^{-\beta\Delta F}$. The curves in Fig.~\ref{fig:excesswork}(b) are obtained by integrating Eq.~\eqref{eq:dotGamma} for each value of $u$.

Figure~\ref{fig:excesswork} shows the scaled excess work $\beta W^\mathrm{ex}$ and generating function $\chi(u)$ for varying frequency $\omega$, with parameters set as $\kappa = 1$, $m = 1$, and different inverse temperatures $\beta = 0.01, 1, 2$. In Fig.~\ref{fig:excesswork}(a), we consider compression processes with $\omega$ varied from $0.2$ to $5$. The scaled excess work is compared between the slow-driving optimal protocol $\omega_\mathrm{op}(s)$ (dots) and exponential protocol $\omega_\mathrm{exp}(s)=\omega_0 (\omega_\tau/\omega_0)^{s}$ (curves). In the slow driving regime, the scaled excess work of the optimal protocol approaches the leading prediction $\beta\mathfrak{L}^2/\tau$, shown by the black dashed lines.  
In Fig.~\ref{fig:excesswork}(b), we show the generating function of work $\chi(u)$ with $\omega$ varied from $1$ to $2$ and the duration $\tau=50$. The work fluctuation theorem \cite{jarzynski_nonequilibrium_1997} is verified at $u=-\beta$, where $\chi(-\beta)=\left\langle e^{-\beta w}\right\rangle =e^{-\beta\Delta F}$. For each temperature, the solid curve gives the propagated generating function $\chi(u)$, while the open circle gives $e^{-\beta\Delta F}$, evaluated independently from the equilibrium free-energy difference. Each circle lies on the corresponding curve at $u/\beta=-1$, providing a direct numerical check of the fluctuation theorem. The moments of work are evaluated from the derivatives at $u=0$. At high temperature ($\beta=0.01$), the quantum results approach the classical limit, consistent with the quantum--classical correspondence principle for work \cite{jarzynski_quantum-classical_2015}.

\subsection{Thermodynamic geometry beyond exact closure}\label{sec:qbm-extensions}

The Brownian model illustrates how the observable dynamics leads directly to thermodynamic geometry. Detailed balance fixes the instantaneous thermal state and hence the equilibrium values of the observables. The closed equations determine the relaxation matrix $L_\lambda$, while $\Theta$ and $\Xi$ describe the control dependence of the Hamiltonian and the equilibrium observables. These quantities give the metric $g_\lambda$, the thermodynamic length, and the slow-driving optimal protocols. The metric is therefore constructed from the observable dynamics rather than assumed in advance.

Exact closure is convenient but not essential. Reference~\cite{chen_boosting_2024} used an operator Krylov expansion to approximate Hamiltonian operator dynamics. Here we adapt the same construction to the adjoint Lindblad dynamics. Starting from the observables coupled to the controls, we repeatedly apply the adjoint generator and project the dynamics onto the finite Krylov space. The size of this space can be increased until the projected dynamics and the corresponding metric converge. As in Ref.~\cite{chen_boosting_2024}, the truncation can be checked from the residual coupling to the next Krylov vectors. The projected relaxation matrix and equilibrium response can then be inserted into the same expression for $g_\lambda$. This gives a practical route to the thermodynamic geometry of interacting systems without evolving the full density matrix.

\section{Conclusion and outlook}\label{sec:conclusion}

We have presented a Lindblad framework for quantum stochastic thermodynamics to study the nonequilibrium thermodynamics of open quantum systems. Each bath dissipator satisfies the local quantum detailed balance condition, ensuring thermodynamic consistency of the complete dynamics. The observable dynamics determines the thermodynamic metric and the associated optimal protocol. For an initially thermal system, thermodynamic consistency leads to the joint fluctuation theorem for quantum work and heat, while the tilted master equation gives the corresponding generating function and moments. A closed observable algebra gives an exact finite set of expectation-value equations in place of the density-matrix dynamics. When exact closure is unavailable, a Krylov expansion provides a systematic finite-dimensional approximation and retains the same route to the thermodynamic metric. 
We apply this framework to quantum Brownian motion \cite{caldeira_quantum_1983, breuer_theory_2002, funo_path_2018}, making minimal refinements to the quantum Brownian motion master equation to ensure the quantum detailed balance condition at all temperatures.

The present problem is complementary to the generator optimization studied in Ref.~\cite{chen_boosting_2024}. There, the kinetic coefficients are chosen to increase the spectral gap and shorten the relaxation time. Here the generator is used to determine the thermodynamic metric, and the control schedule is chosen to reduce excess work. Since the spectral gap sets the relaxation scale in the slow-driving expansion, the two forms of optimization can be combined.
The thermodynamic geometry may also be adapted to annealing schedules \cite{rezakhani_quantum_2009}, optimal controls \cite{boscain_introduction_2021}, and finite-time phase transitions \cite{meibohm_finite-time_2022} when the dynamics has a finite relaxation scale. For interacting many-body systems, variational representations \cite{shi_variational_2018,shi_variational_2020} can be combined with the projected observable dynamics to evaluate the thermodynamic metric without evolving the full density matrix.

\appendix

\section{Derivation of the moments and the joint fluctuation relation}
\label{sec:fcs-derivation}

We first expand Eq.~\eqref{eq:eta_dot} to first order in the counting fields $u$ and $v_k$:
\begin{equation}
\begin{aligned}
\dot{\boldsymbol{\eta}}={}&-i[\boldsymbol h,\boldsymbol{\eta}]
+\sum_{k}\mathcal{D}_{k}[\boldsymbol{\eta}]
+\frac{u}{2}\{\partial_t\boldsymbol h,\boldsymbol{\eta}\}\\
&+\sum_k\frac{v_{k}}{2}\left(
\{\boldsymbol h,\mathcal{D}_{k}[\boldsymbol{\eta}]\}
-\mathcal{D}_{k}[\{\boldsymbol h,\boldsymbol{\eta}\}]\right)+\cdots .
\end{aligned}
\end{equation}
Taking the trace gives
\begin{equation}
\begin{aligned}
\mathrm{tr}(\dot{\boldsymbol{\eta}})={}&u\,\mathrm{tr}\left(\boldsymbol{\eta} \frac{\partial \boldsymbol h}{\partial t}\right)
+\sum_k v_k\,\mathrm{tr}\left(\boldsymbol h \mathcal{D}_k[\boldsymbol{\eta}]\right)\\
\approx{}&u\,\mathrm{tr}\left(\boldsymbol{\rho}\frac{\partial \boldsymbol h}{\partial t}\right)
+\sum_k v_{k}\,\mathrm{tr}\left(\boldsymbol h\mathcal{D}_{k}[\boldsymbol{\rho}]\right).
\end{aligned}
\end{equation}
Thus, to first order, the generating function is
\begin{equation}
\begin{aligned}
\chi(u,\{v_{k}\})
&\approx 1+u\int_{0}^{\tau}dt\,\mathrm{tr}\left(\boldsymbol{\rho}\frac{\partial \boldsymbol h}{\partial t}\right)\\
&\quad+\sum_k v_{k}\int_{0}^{\tau}dt\,\mathrm{tr}\left(\boldsymbol h\mathcal{D}_{k}[\boldsymbol{\rho}]\right).
\end{aligned}
\end{equation}
The constant term follows from the normalized initial condition $\mathrm{tr}\,\boldsymbol{\eta}(0)=1$ and does not affect derivatives at zero counting fields.
Therefore, the first-order derivatives at $u=0$ and $v_k=0$ give the average work and the heat exchanged with the $k$th bath, respectively. Their sum gives the change in internal energy and hence the first law for the ensemble averages.

To derive the integral fluctuation relation, set $u=-\beta_S$ and $v_k=-(\beta_S-\beta_k)$. We verify that $\boldsymbol{\eta}(t)=e^{-\beta_S \boldsymbol h(t)}/Z_\mathrm{i}(\beta_S)$ solves Eq.~\eqref{eq:eta_dot}, where $Z_\mathrm{i}(\beta_S)=\mathrm{tr}[e^{-\beta_S \boldsymbol h(0)}]$.
We suppress the explicit time argument of $\boldsymbol h$ below. Since local detailed balance gives $\mathcal D_k[e^{-\beta_k\boldsymbol h}]=0$, Eq.~\eqref{eq:Dvrho} yields
\begin{align}
\mathcal D_{k,-\beta_S+\beta_k}
[e^{-\beta_S\boldsymbol h}]
={}&e^{-\frac{\beta_S-\beta_k}{2}\boldsymbol h}
\mathcal D_k[e^{-\beta_k\boldsymbol h}]
e^{-\frac{\beta_S-\beta_k}{2}\boldsymbol h}
=0.
\end{align}
The counting term for work follows directly from Eq.~\eqref{eq:Wurho}:
\begin{equation}
\mathcal W_{-\beta_S}[e^{-\beta_S\boldsymbol h}]
=\frac{\partial e^{-\beta_S\boldsymbol h}}{\partial t}.
\end{equation}
The commutator in Eq.~\eqref{eq:eta_dot} also vanishes. Therefore,
\begin{align}
\mathrm{RHS}
=\frac{1}{Z_\mathrm{i}(\beta_S)}
\frac{\partial e^{-\beta_S\boldsymbol h}}{\partial t}
=\dot{\boldsymbol{\eta}}
=\mathrm{LHS}.
\end{align}
Tracing the solution at the end of the evolution yields
\begin{align}
\chi\bigl(-\beta_S,\{-(\beta_S-\beta_k)\}\bigr)
=\frac{Z_\mathrm{f}(\beta_S)}{Z_\mathrm{i}(\beta_S)}
=e^{-\beta_S \Delta F_S}.
\end{align}
This gives Eq.~\eqref{eq:FT_work_heat}. The work and heat fluctuation theorems follow as special cases.

\section{Derivation of the closed second-moment dynamics}
\label{sec:qbm-second-moment-derivation}

Equation~\eqref{eq:qbm-second-moments} follows by applying the adjoint Brownian generator to $x^2$, $(xp+px)/2$, and $p^2$. We separate the Hamiltonian and dissipative terms below. Writing $\Gamma=(\Gamma_{x^2},\Gamma_{xp},\Gamma_{p^2})^{\top}$, the result is
\begin{align}
\dot{\Gamma}=\mathscr{L}_\omega(\Gamma)
=\mathscr{H}_\omega(\Gamma)+\mathscr{D}_\omega(\Gamma),
\end{align}
where the Hamiltonian contribution is
\begin{align}
\mathscr{H}_\omega(\Gamma)&=\left(\begin{array}{ccc}
0 & \frac{2}{m} & 0\\
-m\omega^{2} & 0 & \frac{1}{m}\\
0 & -2m\omega^{2} & 0
\end{array}\right)\Gamma,
\end{align}
and the dissipative contribution is
\begin{align}
\mathscr{D}_\omega(\Gamma)&=\left(\begin{array}{c}
-\kappa[1-\operatorname{sech}(\frac{\beta\omega}{2})]\Gamma_{x^2}
+\frac{\kappa}{2m\omega}\tanh(\frac{\beta\omega}{4})\\
-\kappa\Gamma_{xp}\\
-\kappa[1+\operatorname{sech}(\frac{\beta\omega}{2})]\Gamma_{p^2}
+\frac{\kappa m\omega}{2}\coth(\frac{\beta\omega}{4})
\end{array}\right).
\end{align}
Combining these two contributions gives the matrix $L_\omega$ and the noise vector $f$ in Eq.~\eqref{eq:qbm-second-moments}. Since the dissipator satisfies quantum detailed balance, the steady-state solution is
\begin{align}
\Gamma^{\mathrm{eq}}=\left(\begin{array}{c}
\dfrac{1}{2m\omega}\coth(\dfrac{\beta\omega}{2})\\[4pt]
0\\[4pt]
\dfrac{m\omega}{2}\coth(\dfrac{\beta\omega}{2})
\end{array}\right),
\end{align}
in agreement with the thermal state of the harmonic oscillator.

\section{Counting-field term for quantum Brownian motion}
\label{sec:qbm-counting-derivation}

We give the counting term used in Eq.~\eqref{eq:dotGamma}. The identity
$\partial_t e^{u\boldsymbol h/2}e^{-u\boldsymbol h/2}=\int_0^{u/2}ds\,e^{s\boldsymbol h}(\partial_t\boldsymbol h)e^{-s\boldsymbol h}$,
together with the harmonic-oscillator commutation relations, gives
\begin{equation}
\begin{aligned}
\mathcal W_u[\boldsymbol{\eta}]
={}&\frac{um\omega\dot\omega}{4}\bigg[
(1+s_u)\{x^2,\boldsymbol{\eta}\}
+(1-s_u)\left\{\frac{p^2}{m^2\omega^2},\boldsymbol{\eta}\right\}\\
&\hspace{5.5em}
+ic_u\left[\frac{xp+px}{m\omega},\boldsymbol{\eta}\right]\bigg],
\label{eq:qbm-counting-operator}
\end{aligned}
\end{equation}
where $s_u=\sinh(u\omega)/(u\omega)$ and
$c_u=[1-\cosh(u\omega)]/(u\omega)$. Both are understood by continuity at $u=0$, with $s_0=1$ and $c_0=0$.

For the normalized Gaussian moments defined in Sec.~\ref{sec:qbm-counting}, the Gaussian moment identity used in Ref.~\cite{shi_variational_2018} gives
\begin{align}
\mathscr{W}_u(\Gamma_u)
=\frac{um\omega\dot\omega}{4}
(F_\chi,F_{x^2},F_{xp},F_{p^2})^{\top},
\label{eq:qbm-counting-vector}
\end{align}
with
\begin{align}
F_\chi={}&2\chi\left[(1+s_u)\Gamma_{x^2}
+\frac{1-s_u}{m^2\omega^2}\Gamma_{p^2}\right],\\
F_{x^2}={}&\frac{1-s_u}{m^2\omega^2}(4\Gamma_{xp}^2-1)
-\frac{4c_u}{m\omega}\Gamma_{x^2}
+4(1+s_u)\Gamma_{x^2}^2,\\
F_{xp}={}&\frac{4(1-s_u)}{m^2\omega^2}\Gamma_{p^2}\Gamma_{xp}
+4(1+s_u)\Gamma_{xp}\Gamma_{x^2},\\
F_{p^2}={}&\frac{4(1-s_u)}{m^2\omega^2}\Gamma_{p^2}^2
+\frac{4c_u}{m\omega}\Gamma_{p^2}
+(1+s_u)(4\Gamma_{xp}^2-1).
\end{align}
The extended physical evolution $\mathscr{L}_\omega$ has a zero first component and acts on the other components as in Eq.~\eqref{eq:qbm-second-moments}. Equations~\eqref{eq:qbm-counting-vector} and \eqref{eq:qbm-second-moments} therefore form the closed tilted dynamics used for Fig.~\ref{fig:excesswork}(b).

\section*{Acknowledgments}
J.F.C. thanks Ji-Hui Pei, H. T. Quan, Hui Dong, Tao Shi, and Jordi Tura for the helpful discussion. J.F.C. also acknowledges the support received from the European Union's Horizon Europe research and innovation programme through the ERC StG FINE-TEA-SQUAD (Grant No.~101040729). 

The views and opinions expressed here are solely those of the author and do not necessarily reflect those of the funding institutions. Neither of the funding institutions can be held responsible for them.

\bibliographystyle{unsrt}
\bibliography{regular_article}

\begin{thebibliography}{10}

\bibitem{jarzynski_nonequilibrium_1997}
C.~Jarzynski.
\newblock Nonequilibrium {Equality} for {Free} {Energy} {Differences}.
\newblock {\em Physical Review Letters}, 78(14):2690--2693, April 1997.

\bibitem{crooks_entropy_1999}
Gavin~E. Crooks.
\newblock Entropy production fluctuation theorem and the nonequilibrium work
  relation for free energy differences.
\newblock {\em Physical Review E}, 60(3):2721--2726, September 1999.

\bibitem{seifert_stochastic_2012}
Udo Seifert.
\newblock Stochastic thermodynamics, fluctuation theorems and molecular
  machines.
\newblock {\em Reports on Progress in Physics}, 75(12):126001, December 2012.

\bibitem{crooks_measuring_2007}
Gavin~E. Crooks.
\newblock Measuring {Thermodynamic} {Length}.
\newblock {\em Physical Review Letters}, 99(10):100602, September 2007.

\bibitem{schmiedl_optimal_2007}
Tim Schmiedl and Udo Seifert.
\newblock Optimal {Finite}-{Time} {Processes} {In} {Stochastic}
  {Thermodynamics}.
\newblock {\em Physical Review Letters}, 98(10):108301, March 2007.

\bibitem{blaber_optimal_2023}
Steven Blaber and David~A Sivak.
\newblock Optimal control in stochastic thermodynamics.
\newblock {\em Journal of Physics Communications}, 7(3):033001, March 2023.

\bibitem{soret_thermodynamic_2022}
Ariane Soret, Vasco Cavina, and Massimiliano Esposito.
\newblock Thermodynamic consistency of quantum master equations.
\newblock {\em Physical Review A}, 106(6):062209, December 2022.

\bibitem{ohga_thermodynamic_2023}
Naruo Ohga, Sosuke Ito, and Artemy Kolchinsky.
\newblock Thermodynamic {Bound} on the {Asymmetry} of {Cross}-{Correlations}.
\newblock {\em Physical Review Letters}, 131(7):077101, August 2023.

\bibitem{kolchinsky_thermodynamic_2024}
Artemy Kolchinsky, Naruo Ohga, and Sosuke Ito.
\newblock Thermodynamic bound on spectral perturbations, with applications to
  oscillations and relaxation dynamics.
\newblock {\em Physical Review Research}, 6(1):013082, January 2024.

\bibitem{liang_thermodynamic_2024}
Shiling Liang, Paolo De~Los~Rios, and Daniel~Maria Busiello.
\newblock Thermodynamic {Bounds} on {Symmetry} {Breaking} in {Linear} and
  {Catalytic} {Biochemical} {Systems}.
\newblock {\em Physical Review Letters}, 132(22):228402, May 2024.

\bibitem{owen_universal_2020}
Jeremy~A. Owen, Todd~R. Gingrich, and Jordan~M. Horowitz.
\newblock Universal {Thermodynamic} {Bounds} on {Nonequilibrium} {Response}
  with {Biochemical} {Applications}.
\newblock {\em Physical Review X}, 10(1):011066, March 2020.

\bibitem{maes_frenetic_2017}
Christian Maes.
\newblock Frenetic {Bounds} on the {Entropy} {Production}.
\newblock {\em Physical Review Letters}, 119(16):160601, October 2017.

\bibitem{barato_thermodynamic_2015}
Andre~C. Barato and Udo Seifert.
\newblock Thermodynamic {Uncertainty} {Relation} for {Biomolecular}
  {Processes}.
\newblock {\em Physical Review Letters}, 114(15):158101, April 2015.

\bibitem{horowitz_thermodynamic_2020}
Jordan~M. Horowitz and Todd~R. Gingrich.
\newblock Thermodynamic uncertainty relations constrain non-equilibrium
  fluctuations.
\newblock {\em Nature Physics}, 16(1):15--20, January 2020.

\bibitem{timpanaro_thermodynamic_2019}
André~M. Timpanaro, Giacomo Guarnieri, John Goold, and Gabriel~T. Landi.
\newblock Thermodynamic {Uncertainty} {Relations} from {Exchange} {Fluctuation}
  {Theorems}.
\newblock {\em Physical Review Letters}, 123(9):090604, August 2019.

\bibitem{hasegawa_fluctuation_2019}
Yoshihiko Hasegawa and Tan Van~Vu.
\newblock Fluctuation {Theorem} {Uncertainty} {Relation}.
\newblock {\em Physical Review Letters}, 123(11):110602, September 2019.

\bibitem{van_vu_thermodynamic_2023}
Tan Van~Vu and Keiji Saito.
\newblock Thermodynamic {Unification} of {Optimal} {Transport}: {Thermodynamic}
  {Uncertainty} {Relation}, {Minimum} {Dissipation}, and {Thermodynamic}
  {Speed} {Limits}.
\newblock {\em Physical Review X}, 13(1):011013, February 2023.

\bibitem{ma_universal_2018}
Yu-Han Ma, Dazhi Xu, Hui Dong, and Chang-Pu Sun.
\newblock Universal constraint for efficiency and power of a low-dissipation
  heat engine.
\newblock {\em Physical Review E}, 98(4):042112, October 2018.

\bibitem{liang_minimal_2025}
Shiling Liang, Yu-Han Ma, Daniel~Maria Busiello, and Paolo De~Los~Rios.
\newblock Minimal {Model} for {Carnot} {Efficiency} at {Maximum} {Power}.
\newblock {\em Physical Review Letters}, 134(2):027101, January 2025.

\bibitem{zhai_power-efficiency_2025}
R.~X. Zhai and Hui Dong.
\newblock Power-efficiency constraint for chemical motors.
\newblock {\em Physical Review E}, 111(2):024404, February 2025.

\bibitem{talkner_fluctuation_2007}
Peter Talkner, Eric Lutz, and Peter Hänggi.
\newblock Fluctuation theorems: {Work} is not an observable.
\newblock {\em Physical Review E}, 75(5):050102, May 2007.

\bibitem{esposito_nonequilibrium_2009}
Massimiliano Esposito, Upendra Harbola, and Shaul Mukamel.
\newblock Nonequilibrium fluctuations, fluctuation theorems, and counting
  statistics in quantum systems.
\newblock {\em Reviews of Modern Physics}, 81(4):1665--1702, December 2009.

\bibitem{campisi_colloquium_2011}
Michele Campisi, Peter Hänggi, and Peter Talkner.
\newblock \textit{{Colloquium}} : {Quantum} fluctuation relations:
  {Foundations} and applications.
\newblock {\em Reviews of Modern Physics}, 83(3):771--791, July 2011.

\bibitem{strasberg_quantum_2021}
P.~Strasberg.
\newblock {\em Quantum {Stochastic} {Thermodynamics}: {Foundations} and
  {Selected} {Applications}}.
\newblock Oxford University Press, Oxford, 2021.

\bibitem{perarnau-llobet_no-go_2017}
Martí Perarnau-Llobet, Elisa Bäumer, Karen~V. Hovhannisyan, Marcus Huber, and
  Antonio Acin.
\newblock No-{Go} {Theorem} for the {Characterization} of {Work} {Fluctuations}
  in {Coherent} {Quantum} {Systems}.
\newblock {\em Physical Review Letters}, 118(7):070601, February 2017.

\bibitem{hovhannisyan_energy_2024}
Karen~V. Hovhannisyan and Alberto Imparato.
\newblock Energy conservation and fluctuation theorem are incompatible for
  quantum work.
\newblock {\em Quantum}, 8:1336, May 2024.

\bibitem{gherardini_quasiprobabilities_2024}
Stefano Gherardini and Gabriele De~Chiara.
\newblock Quasiprobabilities in {Quantum} {Thermodynamics} and {Many}-{Body}
  {Systems}.
\newblock {\em PRX Quantum}, 5(3):030201, September 2024.

\bibitem{francica_class_2022}
Gianluca Francica.
\newblock Class of quasiprobability distributions of work with initial quantum
  coherence.
\newblock {\em Physical Review E}, 105(1):014101, January 2022.

\bibitem{solinas_full_2015}
P.~Solinas and S.~Gasparinetti.
\newblock Full distribution of work done on a quantum system for arbitrary
  initial states.
\newblock {\em Physical Review E}, 92(4):042150, October 2015.

\bibitem{pei_exploring_2023}
Ji-Hui Pei, Jin-Fu Chen, and H.~T. Quan.
\newblock Exploring quasiprobability approaches to quantum work in the presence
  of initial coherence: {Advantages} of the {Margenau}-{Hill} distribution.
\newblock {\em Physical Review E}, 108(5):054109, November 2023.

\bibitem{upadhyaya_non-abelian_2024}
Twesh Upadhyaya, William~F. Braasch, Gabriel~T. Landi, and Nicole {Yunger
  Halpern}.
\newblock Non-{Abelian} {Transport} {Distinguishes} {Three} {Usually}
  {Equivalent} {Notions} of {Entropy} {Production}.
\newblock {\em PRX Quantum}, 5(3):030355, September 2024.

\bibitem{arvidsson-shukur_properties_2024}
David R~M Arvidsson-Shukur, William~F Braasch~Jr, Stephan De~Bièvre, Justin
  Dressel, Andrew~N Jordan, Christopher Langrenez, Matteo Lostaglio, Jeff~S
  Lundeen, and Nicole {Yunger Halpern}.
\newblock Properties and applications of the {Kirkwood}–{Dirac} distribution.
\newblock {\em New Journal of Physics}, 26(12):121201, December 2024.

\bibitem{cavina_slow_2017}
Vasco Cavina, Andrea Mari, and Vittorio Giovannetti.
\newblock Slow {Dynamics} and {Thermodynamics} of {Open} {Quantum} {Systems}.
\newblock {\em Physical Review Letters}, 119(5):050601, August 2017.

\bibitem{chetrite_quantum_2012}
R.~Chetrite and K.~Mallick.
\newblock Quantum {Fluctuation} {Relations} for the {Lindblad} {Master}
  {Equation}.
\newblock {\em Journal of Statistical Physics}, 148(3):480--501, August 2012.

\bibitem{liu_characteristic_2016}
Fei Liu and Jingyi Xi.
\newblock Characteristic functions based on a quantum jump trajectory.
\newblock {\em Physical Review E}, 94(6):062133, December 2016.

\bibitem{de_chiara_quantum_2022}
Gabriele De~Chiara and Alberto Imparato.
\newblock Quantum fluctuation theorem for dissipative processes.
\newblock {\em Physical Review Research}, 4(2):023230, June 2022.

\bibitem{manzano_quantum_2018}
Gonzalo Manzano, Jordan~M. Horowitz, and Juan~M. R. Parrondo.
\newblock Quantum {Fluctuation} {Theorems} for {Arbitrary} {Environments}:
  {Adiabatic} and {Nonadiabatic} {Entropy} {Production}.
\newblock {\em Physical Review X}, 8(3):031037, August 2018.

\bibitem{ptaszynski_thermodynamics_2019}
Krzysztof Ptaszyński and Massimiliano Esposito.
\newblock Thermodynamics of {Quantum} {Information} {Flows}.
\newblock {\em Physical Review Letters}, 122(15):150603, April 2019.

\bibitem{ostilli_thermalization_2017}
Massimo Ostilli and Carlo Presilla.
\newblock Thermalization of noninteracting quantum systems coupled to blackbody
  radiation: A {Lindblad}-based analysis.
\newblock {\em Physical Review A}, 95(6):062112, June 2017.

\bibitem{alicki_detailed_1976}
Robert Alicki.
\newblock On the detailed balance condition for non-hamiltonian systems.
\newblock {\em Reports on Mathematical Physics}, 10(2):249--258, October 1976.

\bibitem{carlen_gradient_2017}
Eric~A. Carlen and Jan Maas.
\newblock Gradient flow and entropy inequalities for quantum {Markov}
  semigroups with detailed balance.
\newblock {\em Journal of Functional Analysis}, 273(5):1810--1869, September
  2017.

\bibitem{fagnola_generators_2010}
Franco Fagnola and Veronica Umanità.
\newblock Generators of {KMS} {Symmetric} {Markov} {Semigroups} on {B}(h)
  {Symmetry} and {Quantum} {Detailed} {Balance}.
\newblock {\em Communications in Mathematical Physics}, 298(2):523--547,
  September 2010.

\bibitem{ramezani_quantum_2018}
M.~Ramezani, F.~Benatti, R.~Floreanini, S.~Marcantoni, M.~Golshani, and A.~T.
  Rezakhani.
\newblock Quantum detailed balance conditions and fluctuation relations for
  thermalizing quantum dynamics.
\newblock {\em Phys. Rev. E}, 98(5):052104, November 2018.

\bibitem{gilyen_quantum_2024}
András Gilyén, Chi-Fang Chen, Joao~F. Doriguello, and Michael~J. Kastoryano.
\newblock Quantum generalizations of {Glauber} and {Metropolis} dynamics, May
  2024.
\newblock arXiv:2405.20322 [quant-ph].

\bibitem{chen_efficient_2023}
Chi-Fang Chen, Michael Kastoryano, Fernando G. S.~L. Brandão, and András
  Gilyén.
\newblock Efficient quantum thermal simulation.
\newblock {\em Nature}, 646:561--566, 2025.

\bibitem{chen_boosting_2024}
Jin-Fu Chen, Kshiti~Sneh Rai, Patrick Emonts, Donato Farina, Marcin Płodzień,
  Przemyslaw Grzybowski, Maciej Lewenstein, and Jordi Tura.
\newblock Boosting thermalization of classical and quantum many-body systems.
\newblock {\em Physical Review B}, 113(18):184313, May 2026.

\bibitem{caldeira_quantum_1983}
A.O Caldeira and A.J Leggett.
\newblock Quantum tunnelling in a dissipative system.
\newblock {\em Annals of Physics}, 149(2):374--456, September 1983.

\bibitem{breuer_theory_2002}
Heinz-Peter Breuer and F.~Petruccione.
\newblock {\em The theory of open quantum systems}.
\newblock Oxford University Press, Oxford ; New York, 2002.
\newblock OCLC: ocm49872077.

\bibitem{dechant_underdamped_2017}
A.~Dechant, N.~Kiesel, and E.~Lutz.
\newblock Underdamped stochastic heat engine at maximum efficiency.
\newblock {\em EPL (Europhysics Letters)}, 119(5):50003, September 2017.

\bibitem{chen_microscopic_2022}
Y.~H. Chen, Jin-Fu Chen, Zhaoyu Fei, and H.~T. Quan.
\newblock Microscopic theory of the {Curzon}-{Ahlborn} heat engine based on a
  {Brownian} particle.
\newblock {\em Physical Review E}, 106(2):024105, August 2022.

\bibitem{jarzynski_quantum-classical_2015}
Christopher Jarzynski, H. T. Quan, and Saar Rahav.
\newblock Quantum-{Classical} {Correspondence} {Principle} for {Work}
  {Distributions}.
\newblock {\em Physical Review X}, 5(3):031038, September 2015.

\bibitem{pontryagin1962}
L.~S. Pontryagin, V.~G. Boltyanskii, R.~V. Gamkrelidze, and E.~F. Mishchenko.
\newblock {\em The Mathematical Theory of Optimal Processes}.
\newblock Wiley, New York, 1962.

\bibitem{boscain_introduction_2021}
U.~Boscain, M.~Sigalotti, and D.~Sugny.
\newblock Introduction to the {Pontryagin} {Maximum} {Principle} for {Quantum}
  {Optimal} {Control}.
\newblock {\em PRX Quantum}, 2(3):030203, September 2021.

\bibitem{Cavina2018}
Vasco Cavina, Andrea Mari, Alberto Carlini, and Vittorio Giovannetti.
\newblock Optimal thermodynamic control in open quantum systems.
\newblock {\em Phys. Rev. A}, 98(1):012139, jul 2018.

\bibitem{chen_optimal_2024}
Jin-Fu Chen and H.~T. Quan.
\newblock Optimal control theory for maximum power of {Brownian} heat engines.
\newblock {\em Physical Review E}, 110(4):L042105, October 2024.

\bibitem{sivak_thermodynamic_2012}
David~A. Sivak and Gavin~E. Crooks.
\newblock Thermodynamic {Metrics} and {Optimal} {Paths}.
\newblock {\em Physical Review Letters}, 108(19):190602, May 2012.

\bibitem{scandi_thermodynamic_2019}
Matteo Scandi and Martí Perarnau-Llobet.
\newblock Thermodynamic length in open quantum systems.
\newblock {\em Quantum}, 3:197, October 2019.
\newblock arXiv: 1810.05583.

\bibitem{chen_extrapolating_2021}
Jin-Fu Chen, C.~P. Sun, and Hui Dong.
\newblock Extrapolating the thermodynamic length with finite-time measurements.
\newblock {\em Phys. Rev. E}, 104(3):034117, September 2021.

\bibitem{fei_quantum_2018}
Zhaoyu Fei, H.~T. Quan, and Fei Liu.
\newblock Quantum corrections of work statistics in closed quantum systems.
\newblock {\em Physical Review E}, 98(1):012132, July 2018.

\bibitem{jarzynski_master_1997}
Christopher Jarzynski.
\newblock Equilibrium free-energy differences from nonequilibrium measurements:
  A master-equation approach.
\newblock {\em Physical Review E}, 56(5):5018--5035, 1997.

\bibitem{hummer_free_2001}
Gerhard Hummer and Attila Szabo.
\newblock Free energy reconstruction from nonequilibrium single-molecule
  pulling experiments.
\newblock {\em Proceedings of the National Academy of Sciences},
  98(7):3658--3661, 2001.

\bibitem{jarzynski_classical_2004}
Christopher Jarzynski and Daniel~K. Wójcik.
\newblock Classical and {Quantum} {Fluctuation} {Theorems} for {Heat}
  {Exchange}.
\newblock {\em Physical Review Letters}, 92(23):230602, June 2004.

\bibitem{alicki_quantum_1979}
R~Alicki.
\newblock The quantum open system as a model of the heat engine.
\newblock {\em Journal of Physics A: Mathematical and General},
  12(5):L103--L107, May 1979.

\bibitem{quan_quantum_2007}
H.~T. Quan, Yu-xi Liu, C.~P. Sun, and Franco Nori.
\newblock Quantum thermodynamic cycles and quantum heat engines.
\newblock {\em Physical Review E}, 76(3):031105, September 2007.

\bibitem{funo_path_2018}
Ken Funo and H. T. Quan.
\newblock Path {Integral} {Approach} to {Quantum} {Thermodynamics}.
\newblock {\em Physical Review Letters}, 121(4):040602, July 2018.

\bibitem{chen_quantumclassical_2021}
Jin-Fu Chen, Tian Qiu, and Hai-Tao Quan.
\newblock Quantum–{Classical} {Correspondence} {Principle} for {Heat}
  {Distribution} in {Quantum} {Brownian} {Motion}.
\newblock {\em Entropy}, 23(12):1602, November 2021.

\bibitem{rezakhani_quantum_2009}
A.~T. Rezakhani, W.-J. Kuo, A.~Hamma, D.~A. Lidar, and P.~Zanardi.
\newblock Quantum {Adiabatic} {Brachistochrone}.
\newblock {\em Phys. Rev. Lett.}, 103(8):080502, August 2009.

\bibitem{meibohm_finite-time_2022}
Jan Meibohm and Massimiliano Esposito.
\newblock Finite-{Time} {Dynamical} {Phase} {Transition} in {Nonequilibrium}
  {Relaxation}.
\newblock {\em Physical Review Letters}, 128(11):110603, March 2022.

\bibitem{shi_variational_2018}
Tao Shi, Eugene Demler, and J.~Ignacio~Cirac.
\newblock Variational study of fermionic and bosonic systems with
  non-{Gaussian} states: {Theory} and applications.
\newblock {\em Annals of Physics}, 390:245--302, March 2018.

\bibitem{shi_variational_2020}
Tao Shi, Eugene Demler, and J.~Ignacio Cirac.
\newblock Variational {Approach} for {Many}-{Body} {Systems} at {Finite}
  {Temperature}.
\newblock {\em Physical Review Letters}, 125(18):180602, October 2020.

\end{thebibliography}

\end{document}